\documentclass[fp,twocolumn]{jpsj3}
\usepackage{txfonts}
\usepackage{footmisc}
\usepackage{comment}
\usepackage{color} 
\usepackage{bm}

\title{
First-Principles Study on Cathode Properties of Li$_2M$TiO$_4$ and Na$_2M$TiO$_4$  \\
($M$ = V, Cr, Mn, Fe, Co, Ni)}

\author{Kunihiko Yamauchi$^{1,2}$,  
Hiroyoshi Momida$^{1,2}$ and Tamio Oguchi$^{1,2,3}$}
\inst{
$^1$Elements Strategy Initiative for Catalysts and Batteries (ESICB), Kyoto University, Kyoto 615-8245, Japan \\
$^2$Institute of Scientific and Industrial Research (ISIR-SANKEN), Osaka University, 8-1 Mihogaoka, Ibaraki, Osaka 567-0047, Japan \\
$^3$Center for Spintronics Research Network, Osaka University, Toyonaka, Osaka 560-8531, Japan
}

\date{\today}

\abst{
The cathode properties of  Na$_2M$TiO$_4$ ($M$: transition-metal element) are investigated by means of density-functional-theory calculations. The stability between the layered structure and the disordered structure are focused
in comparison with the Li$_2M$TiO$_4$  prototypical case. 
It is found that the layered structure is more stable than the disordered structure in Na$_2M$TiO$_4$ while those structure shows the similar stability in Li$_2M$TiO$_4$. 
In layered-structure  Na$_2M$TiO$_4$, the formation enthalpies at the intermediate compounds during charge/discharge reactions are significantly low, leading to the unstable voltage-capacity profiles. 
A machine-learning analysis reveals that the total-energy difference between these structures can be described by a simple function of ionic radii.
} 

\newcommand{\lmto}{Li$_2M$TiO$_4$}
\newcommand{\nmto}{Na$_2M$TiO$_4$}
\newcommand{\lmtox}{Li$_{2-x}M$TiO$_4$}
\newcommand{\nmtox}{Na$_{2-x}M$TiO$_4$}
\newcommand{\lmnto}{Li$_2$MnTiO$_4$}
\newcommand{\lco}{LiCoO$_2$}

\definecolor{green}{rgb}{0.0, 0.5, 0.0}

\definecolor{blue}{rgb}{0.0, 0, 0.8}

\begin{document}
\maketitle

\section{\label{sec:introduction}Introduction}


Nowadays, Li-ion rechargeable batteries composed of \lco\ cathode are widely used in our daily life owing to their high energy densities, high voltages, and high capacities. \cite{tarascon.nat2001, ong.ees2011, nitta.mt2015, urban.npjcm2016,islam.csr2014,kubota.ec2020} Nevertheless, the use of the rare and expensive Li and metals has prevented their applications to the large-scale products, such as electric vehicle. On the other hand, Na-ion batteries have attracted much attention as a next candidate for large-scale energy storage since they are made from cheap, abundant, and sustainable Na ions obtained from the oceans or the crust of the earth. \cite{islam.csr2014,kubota.ec2020,yabuuchi.cr2014,kubota.jes2015,kubota.cm2018} Now the key challenge for the Na-ion batteries is to find good cathode materials having both high energy density and high cyclic performance.

A series of \lmto\ ($M$ = V, Cr, Mn, Fe, Co, and Ni) in cation-disordered rock-salt phases has been experimentally studied as high-capacity cathode materials. \cite{sebastian.2003,prabaharan.2004,kuzma.2009,kuezma.2009,dominko.2011,yang.2012,trocoli.2013,kawano.2013,wang.2014,zhang.2015,kitajou.2016} In the disordered structures in mixed transition metal systems, the presence of at least one $d^0$ species such as Ti$^{4+}$ has been known to stabilize the cation disordering in the rock-salt sublattice, \cite{urban.prl2017,clement.ees2020} and this enables the use of a large range of transition metals. It has been also reported that Li diffusion can be fast in disordered cathode materials, especially in their Li-excess phases, due to the Li-percolation mechanism. \cite{clement.ees2020,lee.sci2014,urban.aem2014} Despite such the advantage of cation-disordered cathodes, there is less knowledge of \nmto\ in disordered phases compared with the Li case.

The majority of $AM$O$_2$ ($A$ = Li, Na) cathode materials is known to crystallize in the so-called layered rock-salt type structure (classified as $\alpha$-NaFeO$_2$ type), which is the well-ordered alternating stacks of $M$ and $A$ layers along the NaCl $\langle111\rangle$ direction. \cite{kubota.ec2020} In the case of $A$ = Li, the inclusion of Ti as $A_2M$TiO$_4$ can contribute to mix the $A$, $M$, and Ti sites, resulting in the formation of the disordered rock-salt phase. \cite{urban.prl2017,clement.ees2020,kubota.ec2020} In the case of $A$ = Na, it has been considered that the layered structure is much stable for a wider variety of $M$, probably because of the larger ionic radius of Na$^+$ than that of Li$^+$. However, such the structural stability of \lmto\ and \nmto\ have not been theoretically studied so far. It is necessary to clarify effects of $M$ and $A$ on the structure stability between the layered and the disordered structures using first-principles calculations with a help of machine-learning analyses. To evaluate electrochemical properties such as voltage--capacity curves of them, the stability should be clarified for $A_{2-x}M$TiO$_4$ in the whole range of $x$. First-principles calculations of \lmto\ have been reported to clarify battery reaction mechanism and electrochemical properties such as voltage--capacity characteristics, \cite{hamaguchi.jpsj2018,hamaguchi.ea2020,hamaguchi.ea2020b} though the structural stability between ordered and disordered phases is not clarified yet. 

In this study, aiming at the battery sustainable development goals, we study the structural stability of a series of lithium transition-metal titanates, \lmto\ and \nmto\ ($M$ = V, Cr, Mn, Fe, Co, and Ni), and investigate the potential of \nmto\ as a next-generation cathode candidate. While it has been known that \lmto\ crystallize in the rock-salt structure with random cation distribution \cite{hamaguchi.jpsj2018}, the crystal structure of \nmto\ has not been known with certainty. In our DFT simulation, 
we prepared three candidate crystal structures and compared their structural stability. 

\section{\label{sec:method}Method}

\begin{figure}[htb]
\begin{center}
\includegraphics[width=8.8cm]{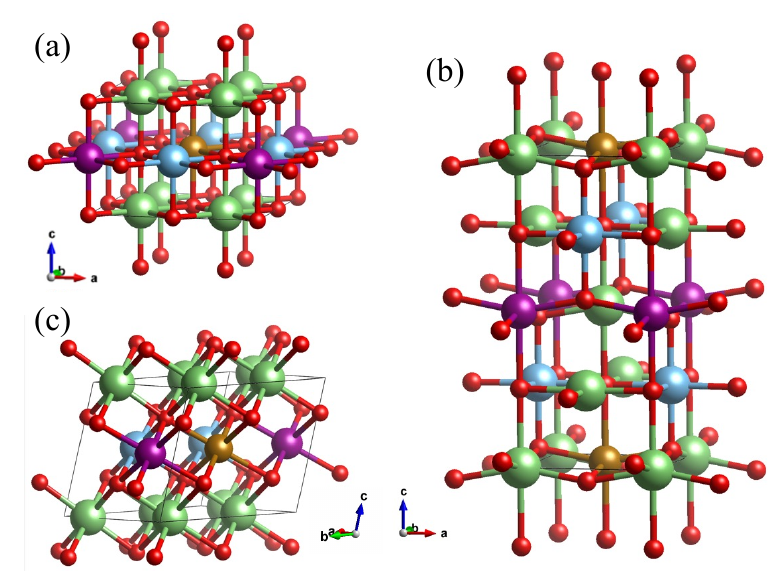}
\caption{\label{fig:crys}
Crystal structure for \lmto\ and \nmto; 
(a) partially disordered ($P4/mmm$),  (b) fully disordered ($I\bar{4}m2$), and (c) layered ($P2/m$) structures.
Green, cyan, purple, brown, red spheres show Li (or Na), Ti, up-spin $M$, down-spin $M$, and O sites, respectively. 
}
\end{center}
\end{figure}

The considered three types of crystal structure for \lmto\ are shown in Fig.\ref{fig:crys}. 
A simple rock-salt structure (called {\it partial disordered structure}, hereinafter) with the space group $P4/mmm$ is the one used in a previous density-functional-theory (DFT) study, showing the pseudo-cubic [001] stacking of the Li (Na) layer and the transition-metal layer.\cite{hamaguchi.jpsj2018} The transition-metal atom and Ti atom are disordered in the same layer. 
A cation {\it fully disordered} rock-salt structure was taken from Materials Project database (mp-754547)\cite{mp}, showing the [001] stacking with a particular tetragonal distortion of oxygen octahedra around Li (Na) and transition-metal atoms. In this structure, all the cation atoms are disordered; Note that those atoms actually show the checkerboard ordering pattern in the super cell, but at least different cation atoms are mixed in the same layer.  
Additionally, a {\it layered} rock-salt structure with $P2/m$ space group was taken from an experimental reference\cite{Lyu.chemmat2015}, showing the [111] stacking.


DFT calculations were performed by using a projector augmented wave method\cite{PAW} implemented in Vienna Ab initio Simulation Package (VASP) code\cite{VASP} by using GGA+$U$ method\cite{PBE,dudarev} with $U$=3 eV for Ti and $M$ 3$d$ orbital states. 
After the atomic structure and the lattice parameters were fully  optimized until forces acting on atoms were smaller than $1 \times 10^{-5}$ eV/\AA, the total energy was calculated self-consistently with the tetrahedron sampling of the $\it{k}$-point mesh of $6 \times 6 \times 8$ for hamaguchi, $6 \times 6 \times 6$ for layered, and $8 \times 8 \times 4$ for disordered structure.

\section{Results and discussion}

\subsection{Structural Stability}

\begin{figure}[htb]
\begin{center}
\includegraphics[width=8cm]{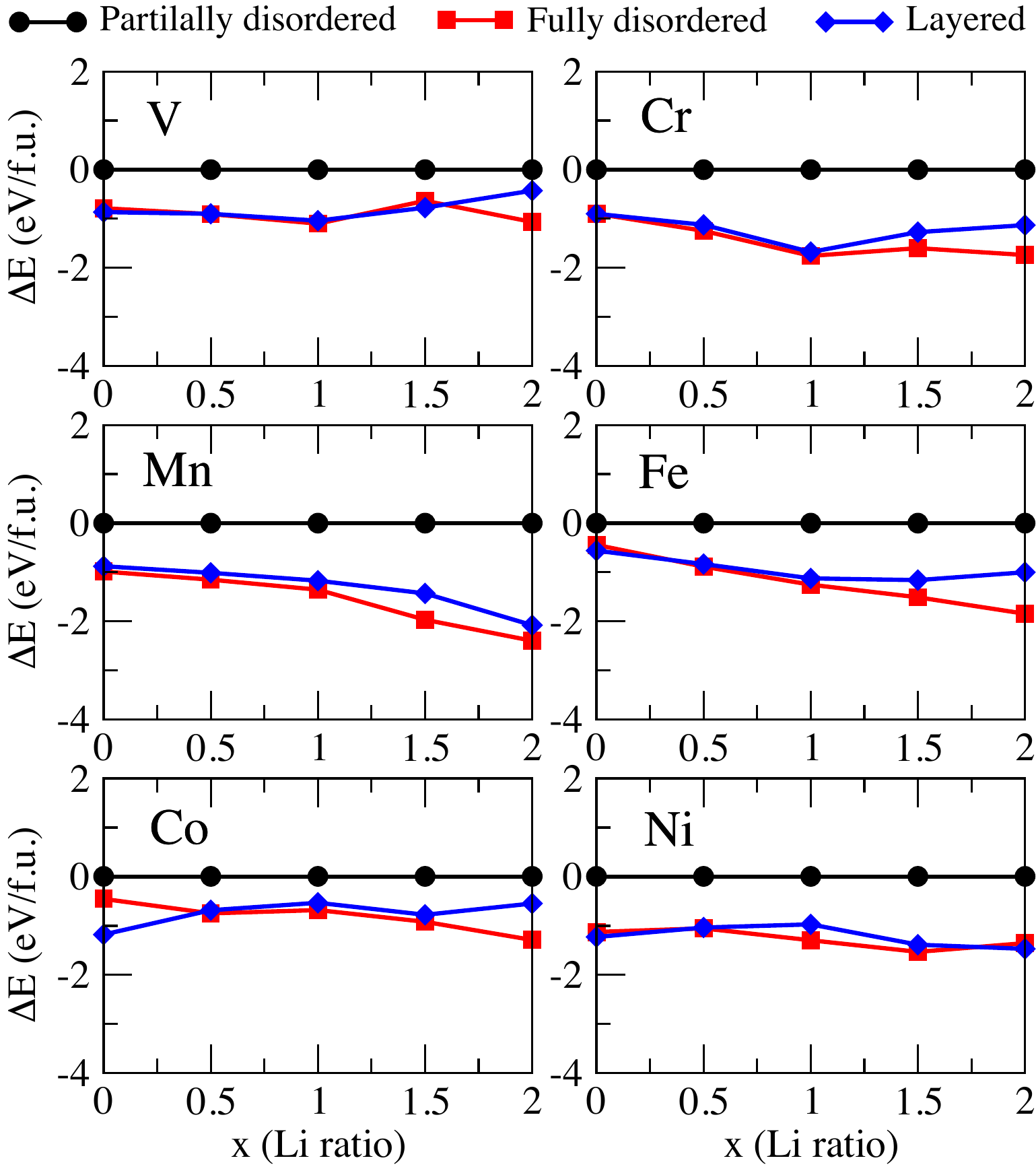}
\\
\includegraphics[width=8cm]{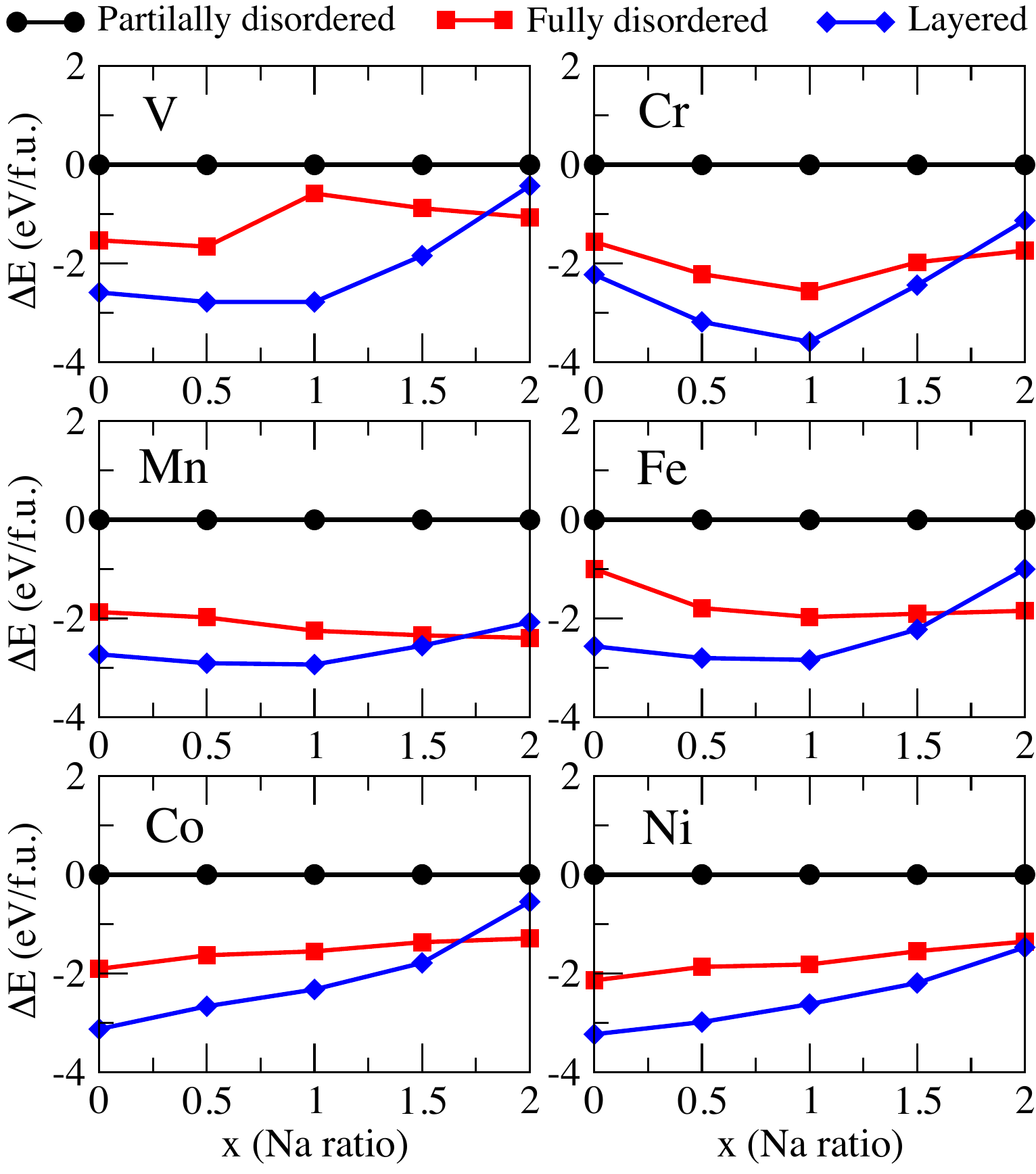}
\caption{\label{fig:ten} 
Total energy comparison between partially disordered, fully disordered, and layered structures for (a) \lmtox\ and  (b) \nmtox. }
\end{center}
\end{figure} 

As shown in Fig. \ref{fig:ten}, the calculated results revealed a general chemical trend:  In \lmto,  the distorted rock-salt structure and the layered structure show the similar stability; 
In \nmto, the layered rock-salt structure is favoured over the distorted rock-salt structure in most cases. This must originate from the difference in the Li and Na ionic size. In section 4, we will come back to the point and discuss  the structural stability based on a machine-learning analysis. 
In the both series, the partially disordered structure used in the previous work\cite{hamaguchi.jpsj2018} is found to be less stable than the other structures. 
In the followings, we will discuss the electronic states, the formation energy, and the expected voltage in the charging process varying the transition-metal element. 

\subsection{Electronic structure}

\begin{figure}[htb]
\begin{center}
\includegraphics[width=8.0cm]{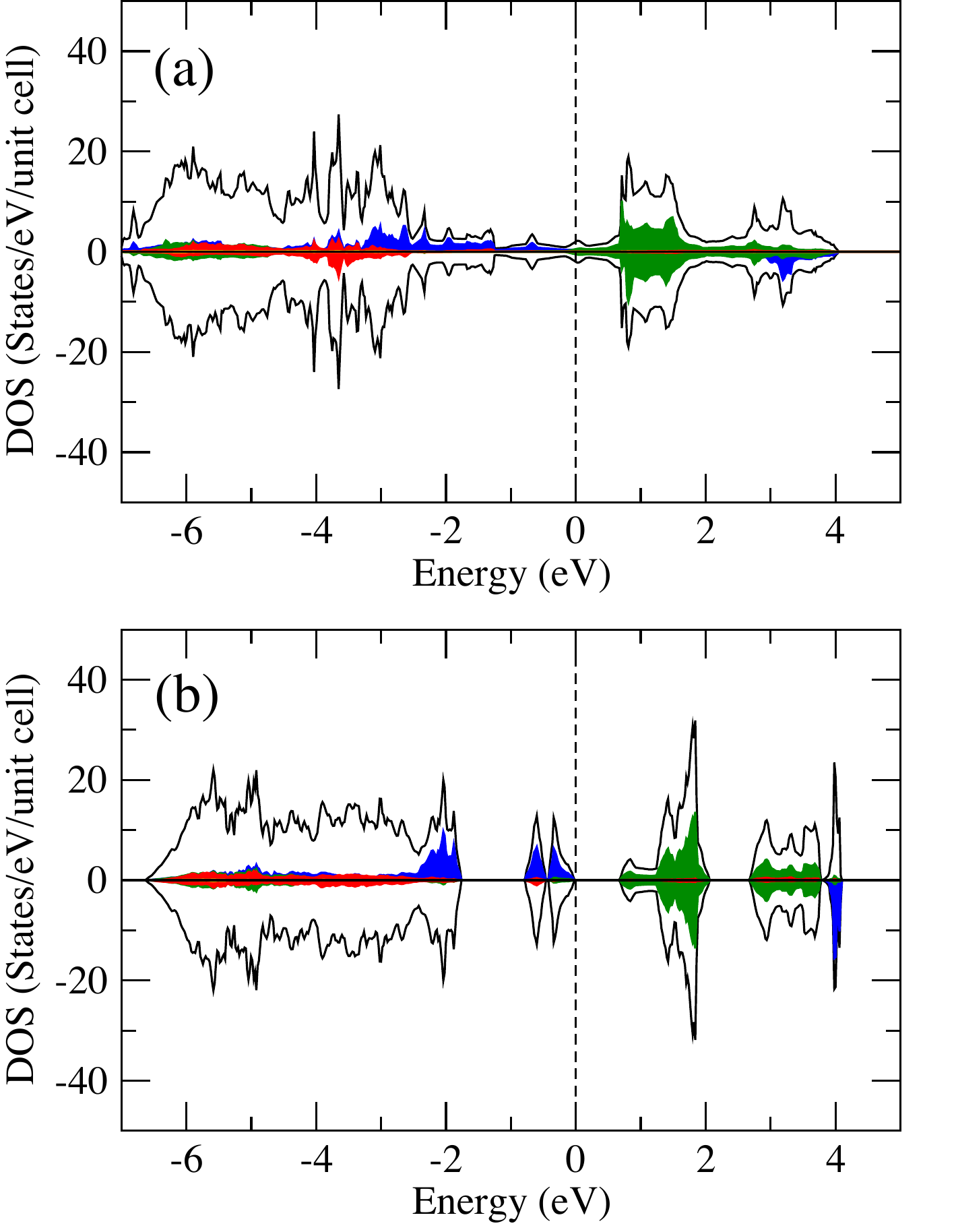}
\caption{\label{fig:dosmn} 
Calculated density of states for (a) fully disordered and (b) layered structures of \lmnto.
Blue, green, red color indicate the Mn-$d$, Ti-$d$, O-$p$ states, respectively. }
\end{center}
\end{figure} 

Figure \ref{fig:dosmn} shows the calculated density of states (DOS) for fully disordered and layered structures of \lmnto. 
In both structures, Mn$^{2+}$ ($d^5$) shows 4.5 $\mu_{\rm B}$  spin moment.  
Li and Na $s$ states lie out of the energy range and they do not affects the states near the Fermi energy. Therefore, the different structural stability between Li- and Na- compounds do not originate from the electronic state, but it is attributed to the mechanical origin, such as the difference of the ionic radii or ionic mass. 
The on-site Coulomb repulsion ($U$) in DFT+$U$ approach allows the band gap open at the layered structure.  On the other hand, the disordered structure shows the semi-metallic behavior. 
A previous DFT study has also reported that the LDA+$U$ calculations on partially disordered \lmnto\ result in metallic state even with larger $U$ values \cite{hamaguchi.jpsj2018}. 
This is due to the fact that Mn-$d$ and Ti-$d$ orbital states show rather strong hybridization forming the three-dimensional network in the disordered structure, while the hybridization is limited in the MnTi layer in layered structure leading to more localized Mn-$d$ orbital state.  
Although in reality, both structures show a random disordering pattern and show the insulating state, it is difficult to simulate in DFT simulations.  

\subsection{Cathode property}

\begin{figure}[htb]
\begin{center}
\includegraphics[width=8.8cm]{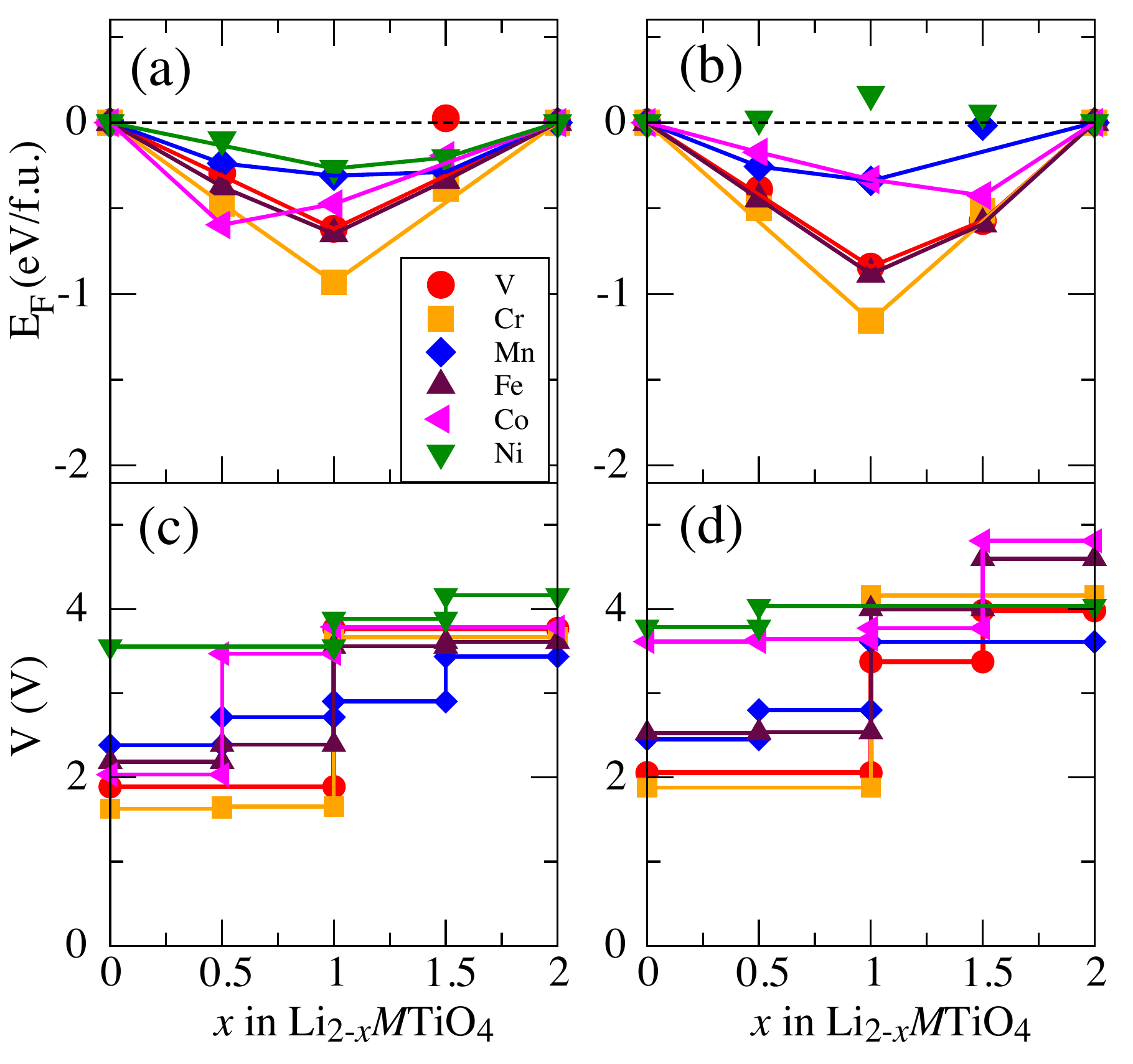}
\caption{\label{fig:volLi} 
Calculated formation enthalpies $E_{\rm F}$ for (a) disordered and (b) layered structures of \lmtox.
Calculated Voltages $V$ for (c) fully disordered and (d) layered structures of \lmtox. }
\end{center}
\end{figure}

\begin{figure}[htb]
\begin{center}
\includegraphics[width=8.8cm]{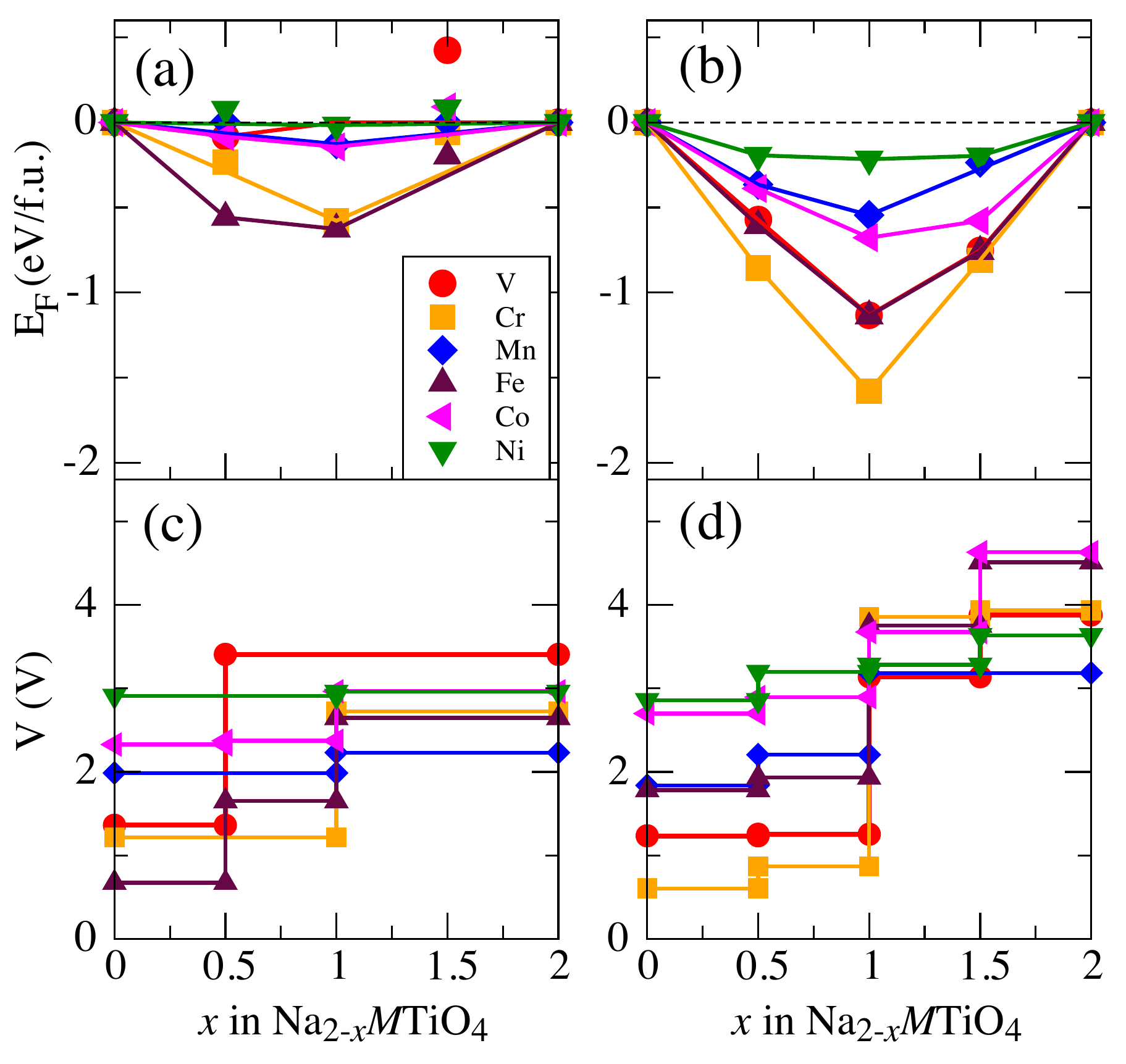}
\caption{\label{fig:volNa} 
Calculated formation enthalpies $E_{\rm F}$ for (a) disordered and (b) layered structures of \nmtox.
Calculated Voltages $V$ for (c) the disordered and (d) the layered structures of \nmtox. }
\end{center}
\end{figure}

Figures \ref{fig:volLi} and \ref{fig:volNa} show the calculated formation enthalpies $E_{\rm F} (x)$ and voltage $V(x)$ for \lmtox\ and \nmtox, respectively, as functions of Li (or Na) reduction $x$. 
Those equations are defined as follows: 
\begin{equation} \label{eq:EF}
    E_{\rm F} (x) =E({\rm Li}_{2-x}M{\rm TiO_4})-\frac{2-x}{2}E({\rm Li_2}M{\rm TiO_4})-\frac{x}{2}E( M{\rm TiO_4})
\end{equation}

\begin{equation} \label{eq:vol}
    V(x)=\frac{E({\rm Li}_{2-x+ \Delta x}M{\rm TiO_4})-E({\rm Li}_{2-x}M{\rm TiO_4})-\Delta x E({\rm Li})}{\Delta x}
\end{equation}
We applied the same equations to Na compounds, too.

\lmtox\ shows the similar trend of $E_{\rm F} (x)$ and $V(x)$ for disordered and layered structures;  
The intermediate compound (with $x$=1) show the deeper formation energy ($-1 \leq E_{\rm F} \leq 0.5 $ eV/f.u.) at $M$=V, Cr, and Fe  and  shallower formation energy ($-0.5 \leq E_{\rm F} \leq +0.2$ eV/f.u.) at $M$=Mn, Co, and Ni. 
The shallowest formation energy at $M$=Ni in the disordered structure results in the high voltage during the changing/discharging process as 3.6 V $\leq V (x) \leq$ 4.1 V showing the almost flat voltage-capacity profile and the best performance among them being consistent with a previous study\cite{hamaguchi.jpsj2018}. 
At $M$=Ni in the layered structure, the discharge reaction does not proceed because of the positive formation energy at the intermediate compounds. 

Making a keen contrast, \nmtox\ shows the very different feature of $E_F(x)$ and $V(x)$ between those two structures. 
Related to the result that the disordered structure is unstable in a range of 0 $\le x \le$ 1.5 in \nmtox\ (see Fig.\ref{fig:ten}),  
the intermediate compounds show either low negative values or positive values. Simultaneously, the voltage shows the steep steps. Among them, only $M$=Mn, Co, and Ni show the rather flat voltage-capacity profile and may be the good candidates for Na-ion-battery cathode materials although the voltage ($V \approx$ 3 V at $M$=Ni) is lower than that of Li compound and it is unsure if the Na reduction reaction proceeds from $x=0$ to $x=2$. The layered structure shows the very deep formation enthalpy for the intermediate compounds and very steep voltage-capacity profile that may be not suitable for the battery applications.

\section{Machine-Learning Analysis}


On closer inspection of Fig.~\ref{fig:ten}, the $M$ and $A$ dependences in the structural stability are found to be roughly determined by the energy difference at $x=0$. Since the $x=0$ phases have all the cation sites fully occupied in the disordered and partially distorted rock-salt structures, as shown in Fig.~\ref{fig:crys}, with the insulating states retained, the stability must be described by an ionic picture, namely the tendency in the ionic radius of $M$ and $A$ cations. To confirm such a postulate, a machine-learning analysis based on the linear regression modeling was performed with empirical ionic radii as the basic descriptors. The target value is the total-energy difference between the fully disordered and layered structures $\Delta E=E[{\rm disorder}]-E[{\rm layer}]$. The effective ionic radii of the $M^{2+}$ and $A^{+}$ cations with coordination number VI, $r(M)$ and $r(A)$, are fetched from data by Shannon and Prewitt\cite{Shannon} as 0.79{\AA} (V$^{2+}$), 0.82{\AA} (Cr$^{2+}$), 0.82{\AA} (Mn$^{2+}$), 0.77{\AA} (Fe$^{2+}$), 0.735{\AA} (Co$^{2+}$), 0.70{\AA} (Ni$^{2+}$), 0.74{\AA} (Li$^{+}$), 1.02{\AA} (Na$^{+}$). In addition, absolute values of the difference between the ionic radii of the $M^{2+}$ and $A^{+}$ cations, $|r(M)-r(A)|$, are included in the descriptors for the regression. For the linear regression modeling, the linearly independent descriptor generation (LIDG) method\cite{fujii, Kanda} was used to generate higher-order descriptors and to detect and remove multicollinearity possibly involved during the descriptor generation. The most appropriate model for the descriptors up to a given order is selected by the cross validation out of the models generated by the exhaustive search method. 


\begin{figure}[htb]
\begin{center}
\includegraphics[width=8.0cm]{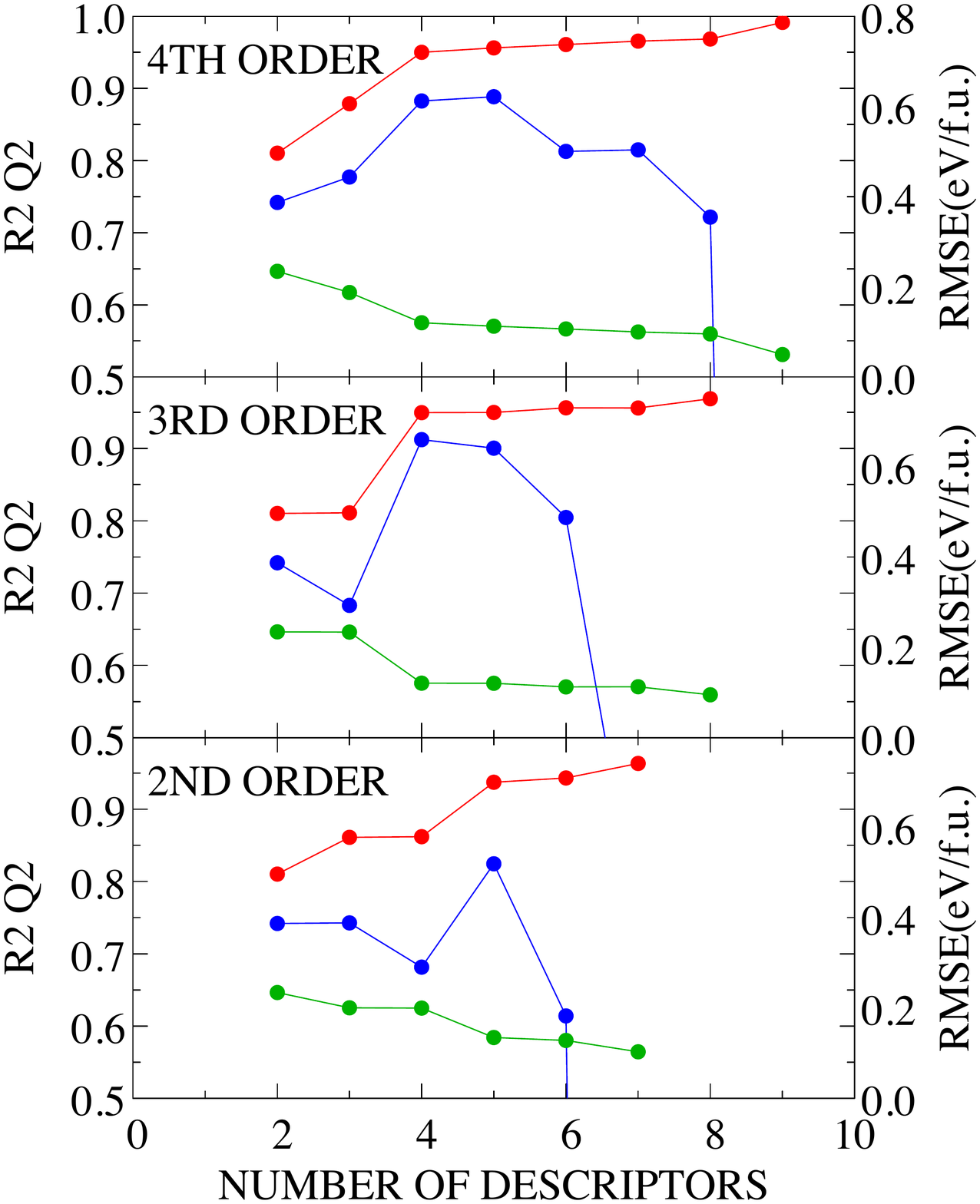}
\caption{\label{fig:R2Q2} 
Decision coefficients $R^2$ (red dots), measures of predictivity $Q^2$ (blue dots), and root mean square error (RMSE) (green dots) obtained by the exhaustive search as a function of the number of descriptors up to $2^{nd}$, $3^{rd}$, and $4^{th}$ orders. The best model for each order is selected by the highest $Q^2$ value. 
}
\end{center}
\end{figure} 

Figure \ref{fig:R2Q2} depicts obtained decision coefficients $R^2$, measures of predictivity $Q^2$, and root mean square error (RMSE) by the exhaustive search as a function of the number of descriptors up to $2^{nd}$, $3^{rd}$, and $4^{th}$ orders. The lager number of descriptors always gives better fitting for the linear regression as indicated by $R^2$, though this often may cause overfitting. The cross validation method can select the best model. For the purpose, the leave-one-out scheme is adopted in the present study and $Q^2$ is used as a measure of predictivity\cite{Kanda}. The best model with the highest $Q^2$ for each order is following. 
\begin{eqnarray}
\Delta E^{(2)} &=& 17.81 - 23.03~r(A) - 199.21~|r(M)-r(A)| \nonumber \\
&+& 92.82~r(M) |r(M)-r(A)|\nonumber \\
&+& 152.47~r(A) |r(M)-r(A)| 
\label{eq:DE2}
\end{eqnarray}
\begin{eqnarray}
\Delta E^{(3)} &=& 0.85 - 25.56~|r(M)-r(A)| \nonumber \\
&+& 377.56~r(M) |r(M)-r(A)|^2 \nonumber \\
&-& 178.44~r(A) |r(M)-r(A)|^2 
\label{eq:DE3}
\end{eqnarray}
\begin{eqnarray}
\Delta E^{(4)} &=& 3.75 - 3.90~r(A) \nonumber \\
&-& 26.17~|r(M)-r(A)| \nonumber \\
&+& 338.69~r(M)^2 |r(M)-r(A)|^2 \nonumber \\
&-& 68.67~r(A)^2 |r(M)-r(A)|^2 
\label{eq:DE4}
\end{eqnarray}
In any model, the ionic radius difference $|r(M)-r(A)|$ is the most important descriptor to govern the structural stability in $A_2M$TiO$_4$ ($A$=Li and Na; $M$=V, Cr, Mn, Fe, Co, and Ni). 

\begin{figure}[htb]
\begin{center}
\includegraphics[width=8.0cm]{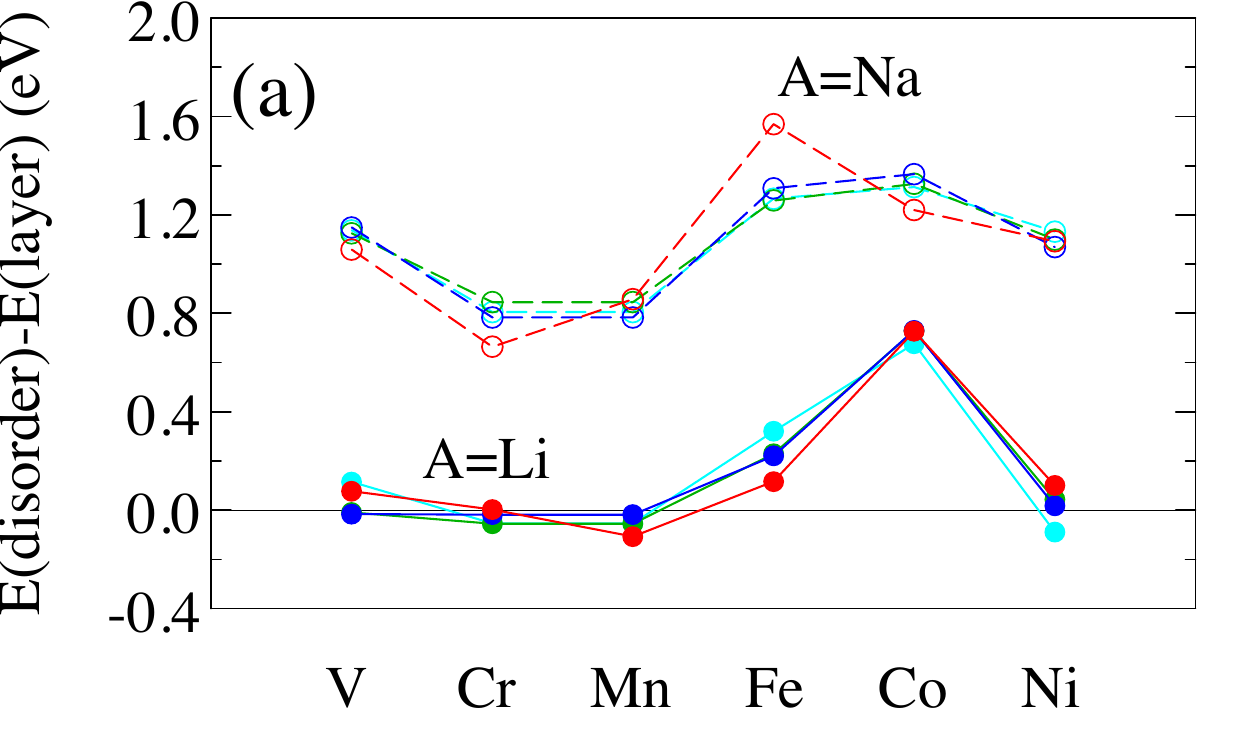}
\includegraphics[width=8.0cm]{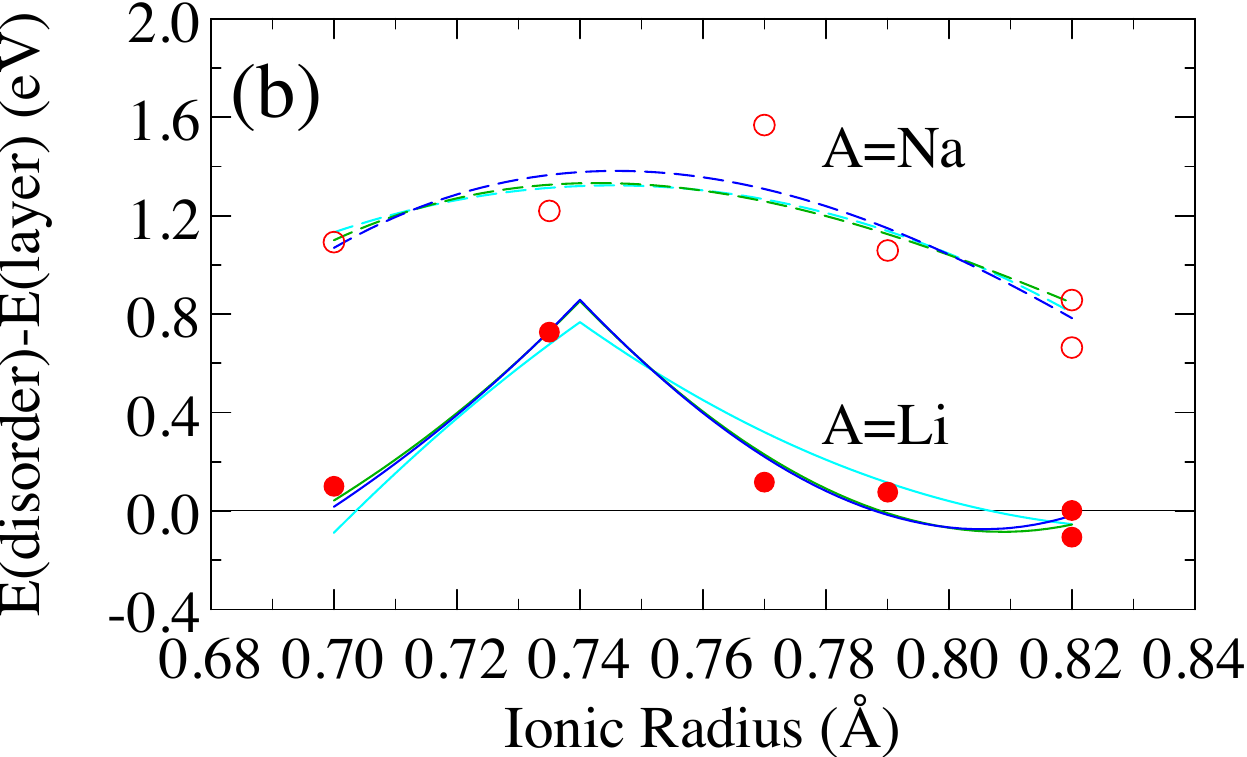}
\caption{\label{fig:E_IR} 
(a) Total-energy difference between the fully disordered and layered structures $\Delta E=E[{\rm disorder}]-E[{\rm layer}]$ by DFT calculation (red dots), LIDG linear regression models with 2$^{nd}$ (cyan dots), 3$^{rd}$ (green dots), and 4$^{th}$ (blue dots) order descriptors in $A_2M$TiO$_4$ ($A$=Li and Na; $M$=V, Cr, Mn, Fe, Co, and Ni). (b) Total-energy difference by the LIDG models (Eqs.~(\ref{eq:DE2}), (\ref{eq:DE3}), and (\ref{eq:DE4})) with 2$^{nd}$ (cyan lines), 3$^{rd}$ (green lines), and 4$^{th}$ (blue dots) order descriptors together with the DFT total-energy difference (red dots for $A$=Li and red circles for $A$=Na) as a function of the ionic radius of $M$ cation. Solid and broken lines are for $A$=Li and Na, respectively.
}
\end{center}
\end{figure} 
Figure \ref{fig:E_IR} (a) shows that the LIDG models well reproduce the DFT total-energy difference for the whole $M$ cations and $A$=Li and Na. Although the model formulae in Eqs.~(\ref{eq:DE2}), (\ref{eq:DE3}), and (\ref{eq:DE4}) look different at a glance, their ionic-radius dependences are almost equivalent as shown in Fig.~\ref{fig:E_IR} (b). In the series of $A$=Li, the total-energy difference has a peak at $M$=Co, implying the significant stability of the layered structure against the fully disordered one. The similar ionic radii of Co$^{2+}$ and Li$^{+}$ in six-coordination should be the key in the layered stability in the cases of $A$=Li. 
On the other hand, in the series of $A$=Na, the layered structure is generally stable throughout the $M$ cations and a minor S-shape variation in $\Delta E$ seen in Fig.~\ref{fig:E_IR} (a) originates from the week ionic-radius dependence in Fig.~\ref{fig:E_IR} (b). The ionic radius of Na$^{+}$ is much larger than that of $M$, leading to the stability of the layered structure. As the Na content is decreased, the relative stability of the layered structure may become gradually weak, as shown in Fig.~\ref{fig:ten}.

\section{Summary}

We investigated the structural stability and the cathode property of \lmto\ and \nmto\ by means of the DFT calculations and the machine learning approach. 
While the disordered structure has an advantage that two Li/Na ions can contribute to the charging reaction, larger Na ion favors the layered structure as the ground state. 
In this study, we found that 
there is a strong relation between the cation ionic radii and the structural stability by using the machine learning method. This knowledge may be useful for the future work to stabilize the disordered structure, for example, by replacing Ti atom by larger Zr atom. 
The present study may lead to the future industrial application to the sodium-ion rechargeable battery. 

\acknowledgment
We acknowledge S. Okada  and A. Kitajou for their helpful comments. 
This work was performed under the research program at the “Dynamic Alliance for Open Innovation Bridging Human, Environment and Materials” in Network Joint Research Center for Materials and Devices and the management of the Elements Strategy Initiative for Catalysts and Batteries (ESICB) supported by the Ministry of Education, Culture, Sports, Science and Technology, Japan (MEXT). 
A part of the computation in this work has been done by using the facilities of the Supercomputer Center, Institute for Solid State Physics, the University of Tokyo. 
The crystallographic figure was generated using VESTA program.\cite{vesta}

\end{document}